\documentclass[11pt]{article}
\usepackage{moriond,epsfig}

\def\Journal#1#2#3#4{{#1} {\bf #2}, #3 (#4)}

\def\PLB{{\em Phys. Lett.}  B}
\def\PRL{\em Phys. Rev. Lett.}
\def\PRD{{\em Phys. Rev.} D}

\def\PRT{\em Phys. Rept.}

\def\be{\begin{equation}}
\def\ee{\end{equation}}
\def\bea{\begin{eqnarray}}
\def\eea{\end{eqnarray}}

\newcommand{\ifm}[1]{\relax\ifmmode #1\else $#1$5\fi}
 
\newcommand{\beq}{\begin{equation}}
\newcommand{\eeq}{\end{equation}}
\newcommand{\beqn}{\begin{eqnarray}}
\newcommand{\eeqn}{\end{eqnarray}}
\newcommand{\bi}{\begin{itemize}}
\newcommand{\ei}{\end{itemize}}
\newcommand{\bd}{\begin{description}}
\newcommand{\ed}{\end{description}}
\newcommand{\bHuge}{\begin{Huge}}
\newcommand{\bhuge}{\begin{huge}}
\newcommand{\bLARGE}{\begin{LARGE}}
\newcommand{\bLarge}{\begin{Large}}
\newcommand{\blarge}{\begin{large}}
\newcommand{\eHuge}{\end{Huge}}
\newcommand{\ehuge}{\end{huge}}
\newcommand{\eLARGE}{\end{LARGE}}
\newcommand{\eLarge}{\end{Large}}
\newcommand{\elarge}{\end{large}}

\def \gtsim    {\relax\ifmmode{\mathrel{\mathpalette\oversim >}}
                  \else{$\mathrel{\mathpalette\oversim >}$}\fi}
\def \ltsim    {\relax\ifmmode{\mathrel{\mathpalette\oversim <}}
                  \else{$\mathrel{\mathpalette\oversim <}$}\fi}
\def\oversim#1#2{\lower4pt\vbox{\baselineskip0pt \lineskip1.5pt
            \ialign{$\mathsurround=0pt#1\hfil##\hfil$\crcr#2\crcr\sim\crcr}}}
\def \dk       {\relax\ifmmode{\rightarrow}\else{$\rightarrow$}4\fi}
\def \to       {\relax\ifmmode{\rightarrow}\else{$\rightarrow$}4\fi}
\def \Dk    {\relax\ifmmode{\Rightarrow}\else{$\Rightarrow$}\fi}
\newcounter{minutes}

\newcommand{\met}{\mbox{${E\!\!\!\!/_{\rm T}}$}}

\def \sp       {\relax\ifmmode{\;}\else{$\;$}\fi}	
\newcommand{\gev}  {\mbox{${\rm GeV}$}}
\newcommand{\gevc} {\mbox{${\rm GeV}/c$}}

\newcommand{\gevcc}{\mbox{${\rm GeV}/c^2$}}

\newcommand{\ipb}{\mbox{${\rm pb}^{-1}$}}

\def\Journal#1#2#3#4{{#1} {\bf #2}, #3 (#4)}
\def \PRL      {Phys. Rev. Lett.~}

\def \PRD      {Phys. Rev. D}

\def \PLB      {Phys. Lett. B}

\def \etal     {$et \; al. \;$}
\begin{document}
\vspace*{4cm}
\title{SEARCHES FOR BEYOND SM HIGGS BOSON AT THE TEVATRON}

\author{A. Safonov \\ (for CDF and D0 Collaborations)}

\address{Department of Physics, Texas A\&M University, 4242 TAMU,\\
College Station, TX 77843, USA}

\maketitle\abstracts{In the following, we describe preliminary 
results of searches for non-SM higgs bosons at the CDF and D0 
experiments. Both experiments use data obtained in $p\bar{p}$ 
collisions at the Tevatron at $\sqrt{s}=1.96$ TeV.}

\section{Introduction}

The Higgs mechanism is the heart of the Standard Model (SM) providing
masses to gauge bosons via electroweak symmetry breaking (EWSB). However,
the SM fails to explain the origin of the Higgs mechanism and has 
certain naturalness problems, such as the extreme fine tuning
required to keep the higher order corrections to the higgs mass
under control (the so called hierarchy problem). In popular SM
extensions, such as minimal SUSY (MSSM)~\cite{mssm} and Little(st) Higgs~\cite{little_higgs}
models, the EWSB is radiatively 
generated and these models are free from many of the above difficulties. 
In these models, the higgs sector is typically more complex, requiring
additional neutral and charged higgs bosons in MSSM, and even 
doubly charged higgs bosons in Little Higgs, SUSY with 
extended higgs sector, and Left-Right symmetric~\cite{lr-models} models. 
While the production cross section of higgs bosons is usually model 
dependent, the phenomenology in terms of decay modes 
and production mechanisms are often similar for a 
wide variety of new phenomena scenarios. This simplifies 
the interpretation of the experimental results: while the
cross section driven mass limits are usually model 
dependent, production cross section limits obtained for
certain benchmark scenarios are often applicable to other 
models.
\section{Neutral Higgs Boson Searches}\label{subsec:prod}

Apart from the SM, neutral higgs bosons appear in almost every scenario
exploring new phenomena. A typical example is the MSSM, where the 
higgs sector consists of three neutral higgs
bosons: two scalars, $h$ and $H$, and a pseudoscalar $A$ (and also
two charged higgs fileds $H^{\pm}$). Compared to SM, there are differences 
in higgs production mechanisms. In MSSM at large $\tan{\beta}$
there is a strong enhancement of the cross-section for the dominant 
production channel $gg \to \phi$ ($\phi = A$ and $h$ or $H$) 
due to additional diagrams with a $b$-quark in the loop.
The cross-section in this case scales as $\tan^2{\beta}$. Similarly, 
there is a strong enhancement of diagrams
with $b$-quarks in the final state coming from the 
proton/antiproton sea or from radiative pair production leading to
additional experimental signatures of higgs produced in association 
with one or two $b$-jets.

\begin{figure}
\centerline{\epsfig{figure=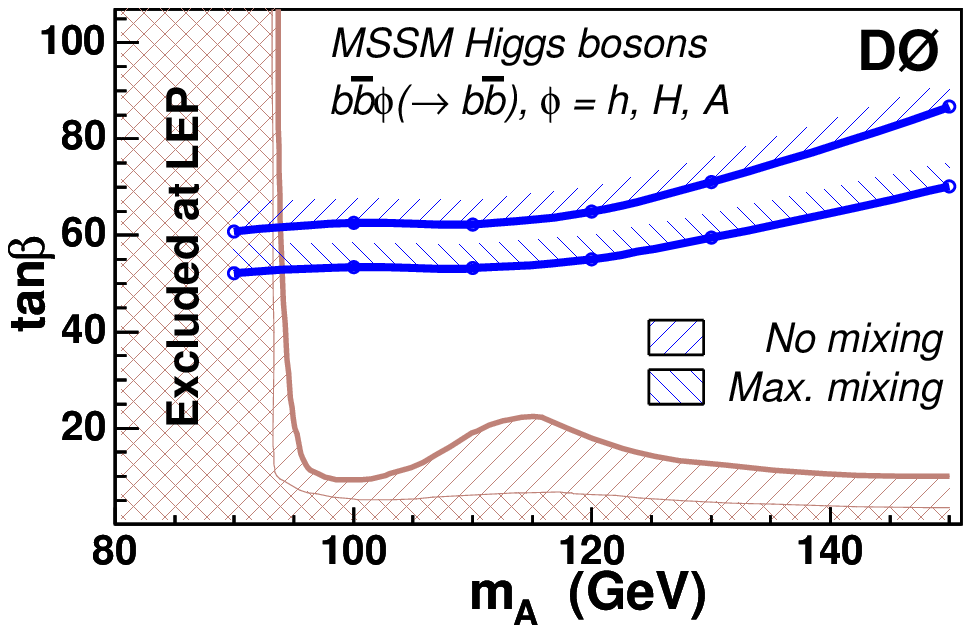,height=1.5in}
\epsfig{figure=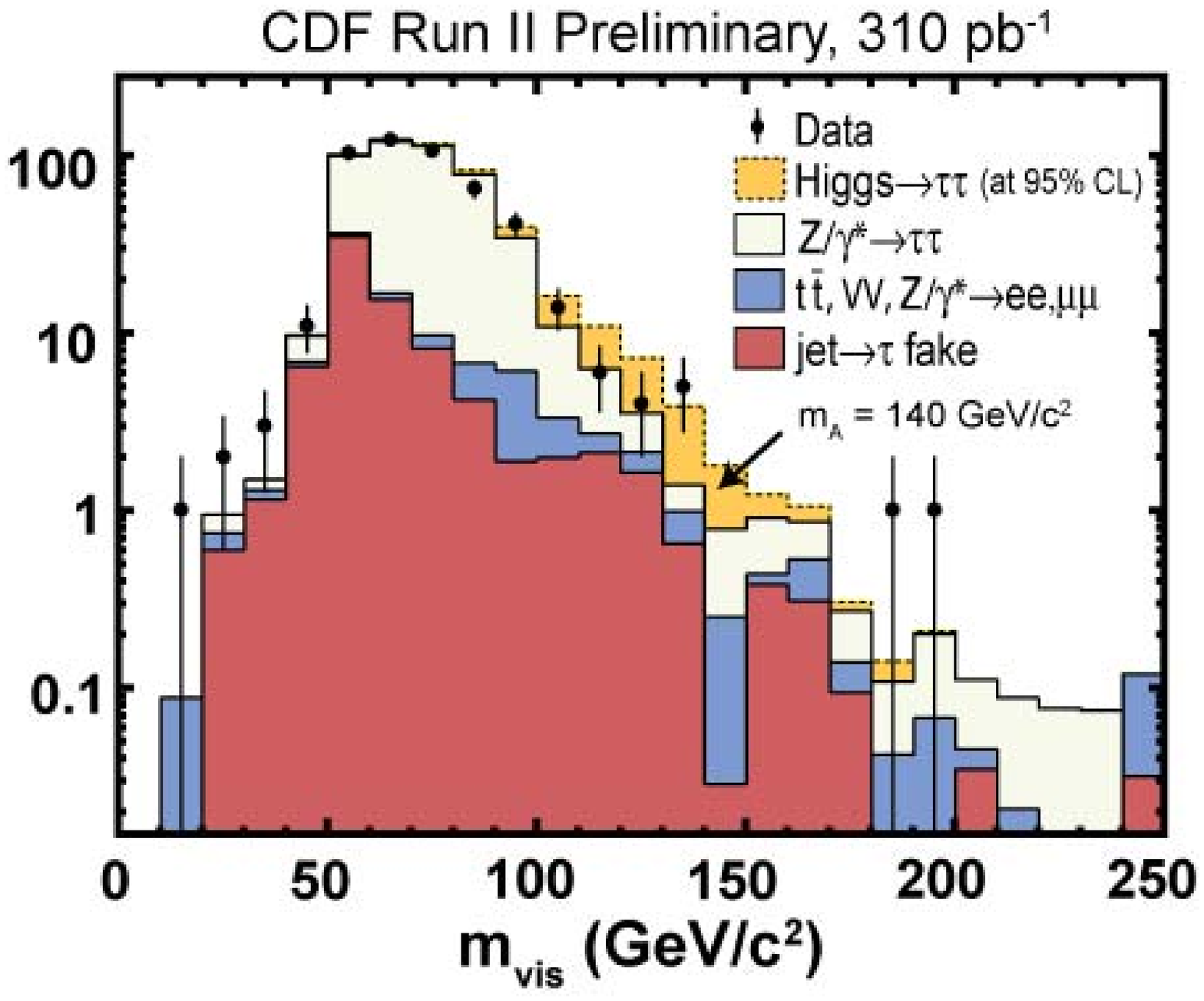,height=1.6in}}
\caption{Left: The exclusion plot in the MSSM $m_A$ versus $\tan{\beta}$ plane for
$hb\bar{b} \to b \bar{b} b \bar{b}$ search. Right: Invariant mass of
lepton, tau and $\met$ in the $h \to \tau \tau$ analysis.
\label{fig:new}}
\end{figure}

For the majority of the MSSM parameter space, light higgs bosons decay 
predominantly to $b\bar{b}$ ($Br\simeq 90$\%) and $\tau \tau$ 
($Br\simeq 8-9$\%). However, one should note that the radiative 
corrections may violate the fermion mass universality of the higgs 
coupling and lead to a strong suppression of the higgs coupling 
to $b$-quarks making taus the best bet for discovering MSSM higgs. 
Experimentally, the use of the $b\bar{b}$ decay modes is most 
effective for processes involving additional objects (e.g. $b$-jets). 
Just as in the SM case, the very high cross section of QCD $b\bar{b}$ 
production at $p \bar{p}$ colliders precludes one from using 
the dominant $gg\to h/A/H$ mode. This loss can be
partially recovered by using the $\tau \tau$ final state that
has a much cleaner signature, but suffers from the 
lower branching ratio compared to $b$-jets. 

\subsection{Search for $Hb(b) \to bbb(b)$}

This D0 analysis\cite{do-hbb} searches for MSSM higgs bosons at large $\tan{\beta}$
decaying to a pair of $b$-jets and produced in association with one
or more additional $b$-jets using 260 $\ipb$ of data. 
Events are required to have at least three jets in the $|\eta| < 2.5$ 
region passing thresholds of 35, 20 and 15 GeV and tagged as $b$-jets 
using the Secondary Vertex (SV) tagging algorithm. The SV tagging 
efficiency is about 55\% for central $b$-jets with $E_T>35$ GeV, 
while the rate of light quark jets being misidentified as $b$-jets 
is $\simeq1$\%. The two highest $E_T$ jets are used to form 
an invariant mass, which is used as the final discriminating variable 
between signal and SM backgrounds. Shape of the backgrounds due to 
mistags is estimated using a data sample selected similar to the signal case 
except that only two jets have to be tagged as $b$-jets (rather than three), 
and where the $E_T$ dependent mistag rates are applied to the non-tagged 
jets.  Simulation shows that the remaining background contributions 
from $b\bar{b}b\bar{b}$ and $Zb\bar{b}\to b\bar{b}b\bar{b}$ have shapes
similar to that of the mistag sample. Data selected in the signal
region is then fitted using shapes of the signal and backgrounds 
(taken from the mistag sample) for the normalization of a possible higgs 
signal. Signal acceptance is obtained using Pythia; production 
cross section and branching ratios come from the CPsuperH program. With 
no deviation from SM, the analysis sets a limit on the production cross 
section of MSSM higgs in this topology. The cross section limit is then 
re-interpreted 
in terms of the MSSM parameters $\tan{\beta}$ and $m_A$ (mass of the 
pseudoscalar). Results are presented in Fig. \ref{fig:new}, lines 
show the part of the parameter space excluded at 95\% C.L.

\subsection{Search for $H \to \tau \tau$}

\begin{figure}
\centerline{\epsfig{figure=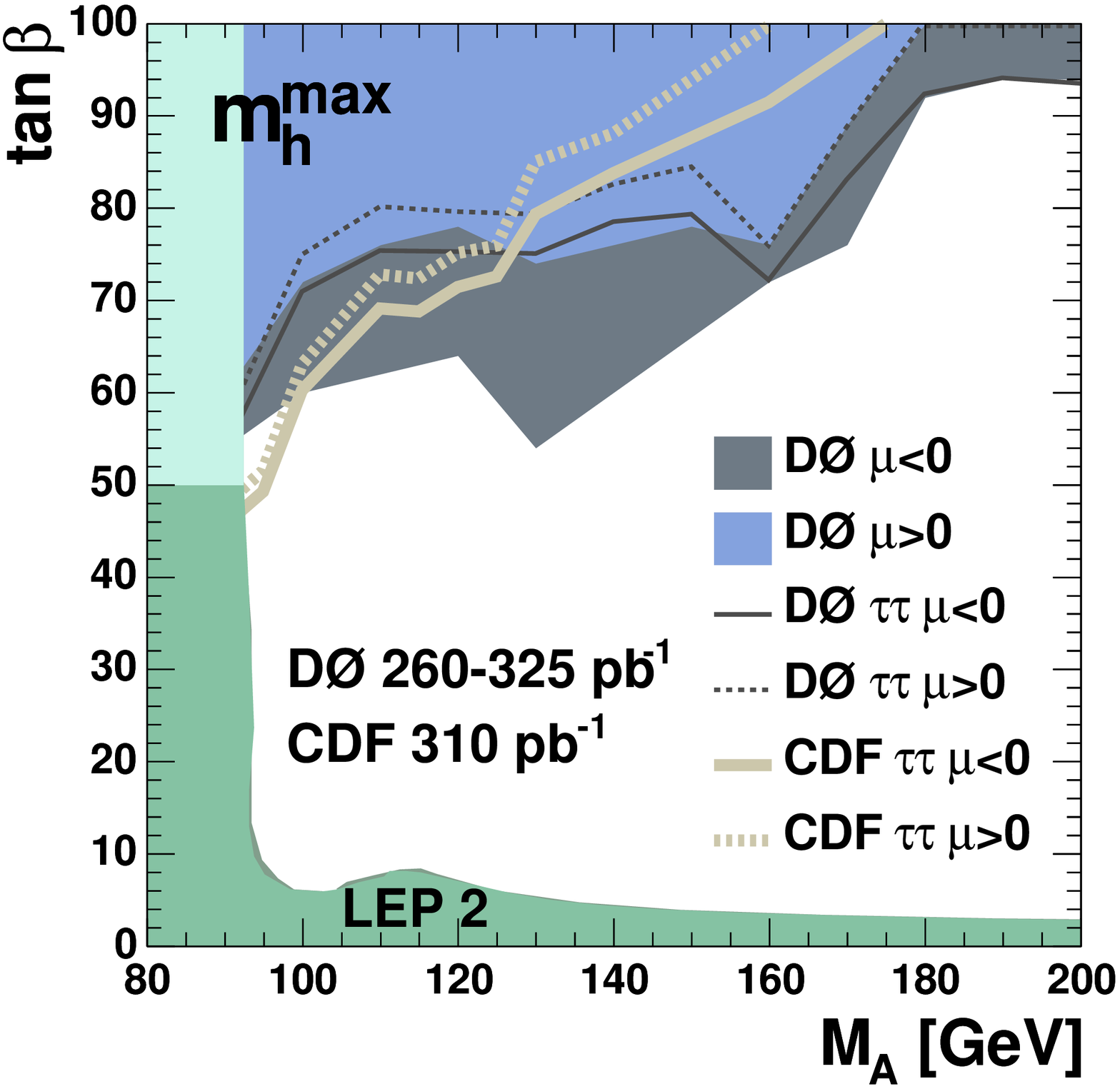,height=1.5in}
\epsfig{figure=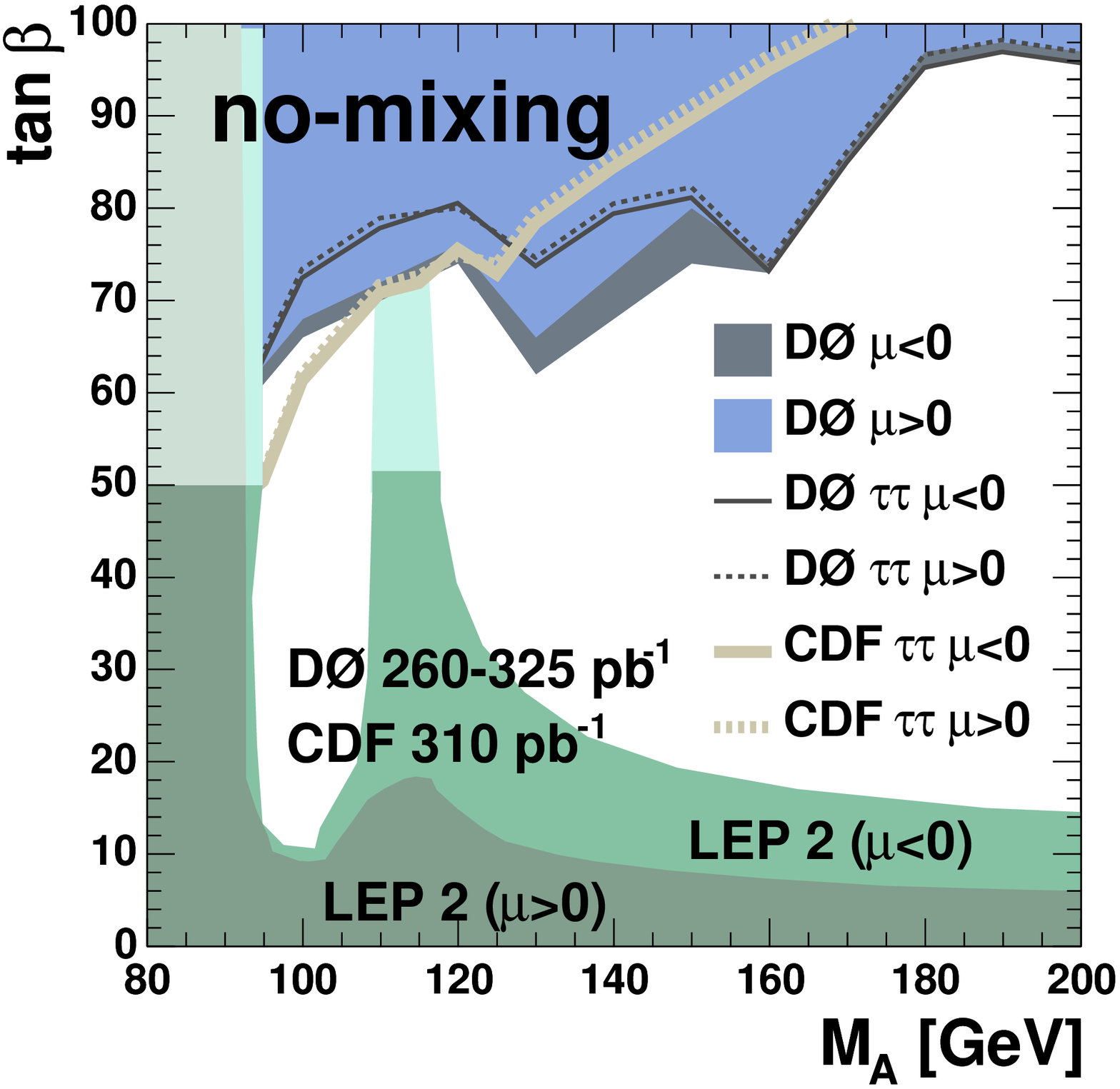,height=1.5in}}
\caption{The exclusion plot in the MSSM $m_A$ versus $\tan{\beta}$ plane for
$H \to \tau \tau$ search for $m_h^{max}$ (left) and no mixing (right) scenarios 
showing both CDF and the combined D0 $H \to \tau \tau$ and $Hb(b) \to bbb(b)$ results.
\label{fig:ditaus}}
\end{figure}

Both CDF and D0 perform a search for higgs decaying to a pair of tau
leptons. The CDF analysis \cite{cdf-tt} uses 320 $\ipb$ and relies on the data
selected using lepton+track triggers. The offline selections require
a reconstructed light lepton ($e$ or $\mu$) with transverse 
energy of 10 GeV and a hadronically decaying tau candidate with
$p_T > 15$ $\gevc$. To 
suppress backgrounds due to QCD jet production, an additional 
requirement $p_T^{\tau}+p_T^{l}+\met >50$ $\gev$ ($\met$ is missing
transverse energy) is applied. Final events are used to construct an 
invariant mass of lepton, tau, and $\met$, and the distribution is fitted 
for a combination of possible signal and SM background (see Fig.~\ref{fig:new}. 
With no
excess found in the data CDF sets a limit on MSSM higgs production 
cross section, which is then re-interpreted in terms of an exclusion 
plot in the $\tan{\beta}$ versus $m_A$ plane shown in Fig. \ref{fig:ditaus}.

The D0 $H \to \tau \tau$ analysis is based on 350 $\ipb$ of data collected 
using inclusive electron and muon triggers as well as an $e+\mu$ trigger. 
Light leptons are required to pass $p_T>14$ $\gevc$. A neural net 
technique is used to select hadronically decaying tau candidates 
with $p_T>20$ $\gevc$. The neural net is separately optimized for
$\pi$, $\rho$ and ``3-prong''-like hadronic tau topologies and 
the main variables allowing discrimination against QCD jets are related 
to isolation and differences in 
shower profile. After additional topological cuts to suppress
backgrounds from multi-jet and $W$+jet production, the distribution
of the invariant mass of lepton, tau, and $\met$ (or two leptons and 
$\met$ for the $e\mu$ channel) is fitted for a combination of 
possible higgs signal and SM backgrounds. With no evidence of the
new physics signal, a 95\% C.L. limit on the maximum alowed cross section
is set. The obtained limit is then interpreted as an exclusion region 
in the plane of $\tan{\beta}$ versus $m_A$ for the same benchmark points
as in the CDF analysis using the FEYNHIGGS program. Figure \ref{fig:ditaus} 
shows the combined limit from this analysis and the one in the 
$Hb(\bar{b}) \to b\bar{b}b(\bar{b})$ channel.

\section{Charged Higgs Searches}

The charged higgs is predicted in the MSSM and appears in other beyond-SM scenarios
with expanded higgs sector. In the MSSM,
direct production of $H^{\pm}$ at the Tevatron is neglible, however,
charged higgs bosons can be produced in top quark decays,
predominantly via $t \to b H^{\pm}$. This new decay mode will modify 
the branching ratios for the top quark decay modes, i.e. lower fraction
of $t \to Wb$. In turn, the $H^{\pm}$ 
decay modes are different from those of $W$ bosons: MSSM 
$H^+$ branching ratios are a strong function of $\tan{\beta}$, but at high
$\tan{\beta}$ $H^+$ preferably decays into $\tau \nu$, while $W$ leptonic decay 
modes are nearly democratic among generations). Therefore, one can 
detect the presence of $H^{\pm}$ by observing a violation of 
the top branching ratios prescribed by the SM. This search~\cite{cdf-h+} uses
192 $\ipb$ of CDF data and analyzes the following top samples: (i) dilepton
($e$ or $\mu$) plus jets; (ii) lepton+jets with exactly one $b$-tag;
(iii) lepton+jets with exactly two or more $b$-tags; and (iv) lepton+hadronic
tau with two or more jets. Signals of new physics will lead to a deviation
of the observed number of events in these samples compared to the SM prediction.
The data is selected using standard selections with lepton and tau $p_T>20$
$\gevc$ and the $b$-jets are tagged using a secondary vertex algorithm.
The number of expected(observed) events in each of the four categories
above is 11(13), 54(49), 10(8) and 2(2). With no disagreement, a limit
is set on the production cross-section. Using MSSM as a reference model, 
the results are interpreted as an exclusion region in the $m_{H^{\pm}}$ 
versus $\tan{\beta}$ plane for several benchmark points. Figure 
\ref{fig:charged} shows the excluded region for one of the selected
points.
%

\section{Doubly Charged Higgs Searches}

\begin{figure}
\centerline{\epsfig{figure=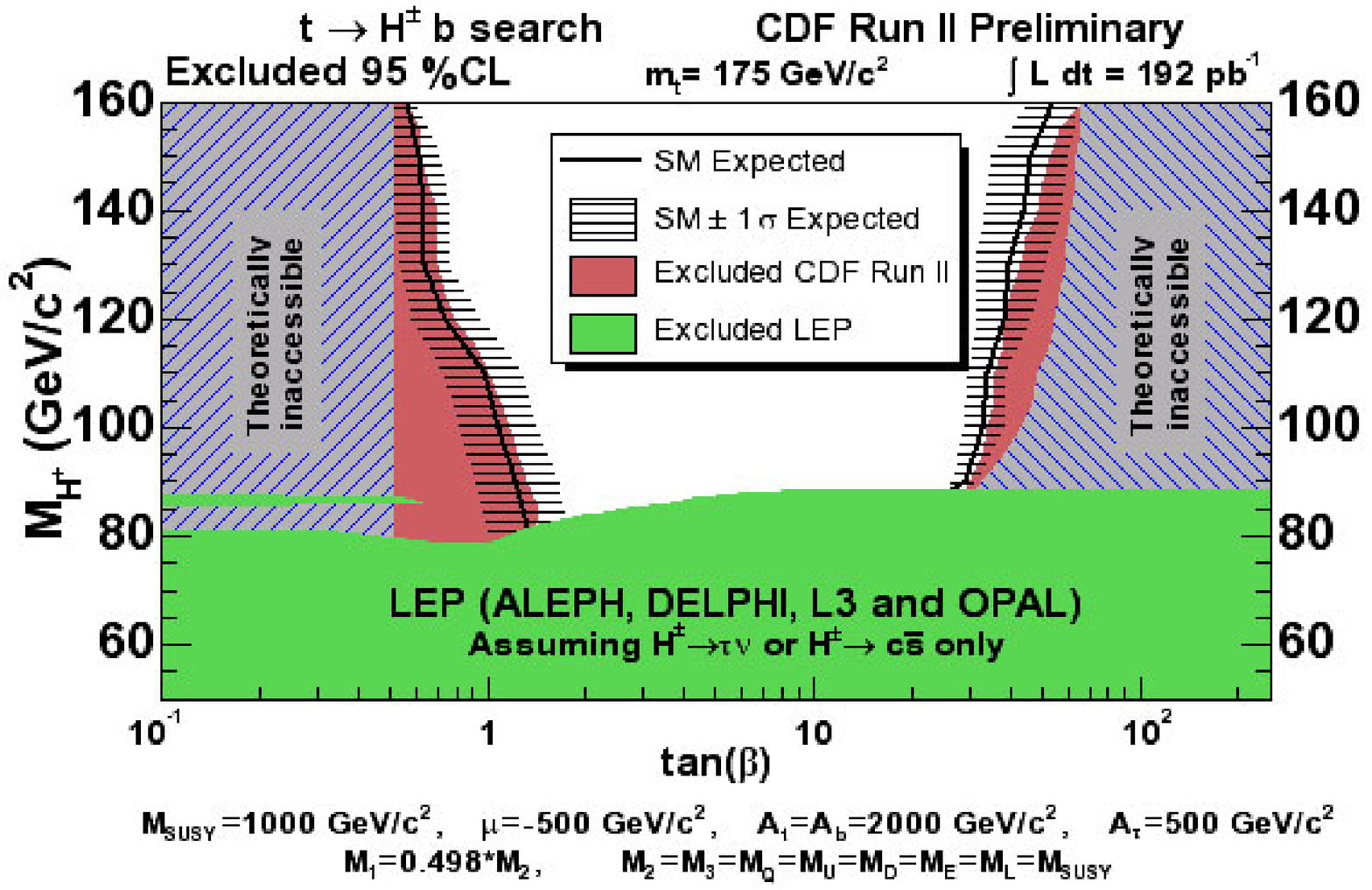,height=1.5in}
\epsfig{figure=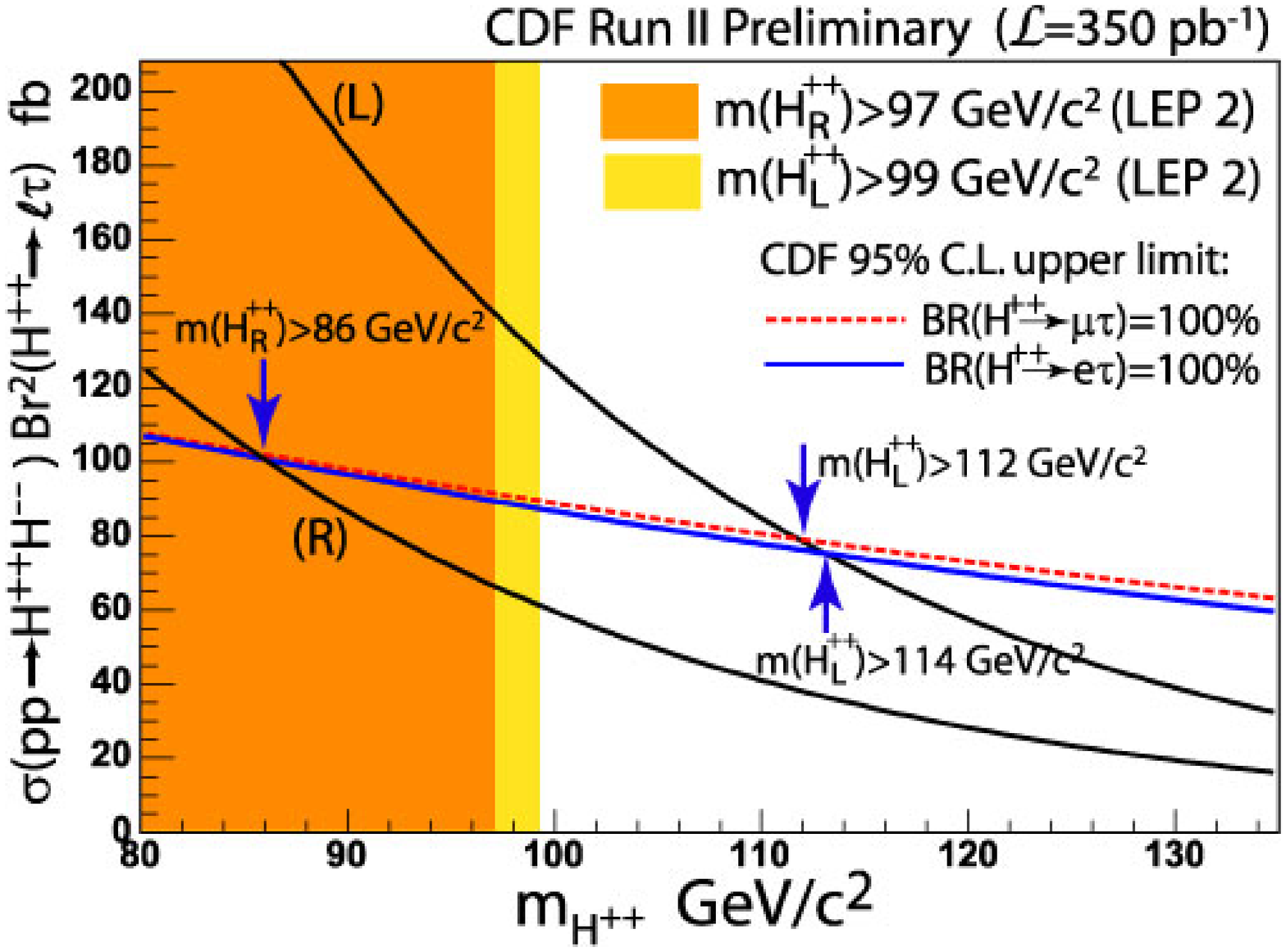,height=1.5in}}
\caption{Left: Exclusion plot in the MSSM $m_{H^{\pm}}$ versus $\tan{\beta}$ plane 
for charged higgs search. Right: 95\% C.L. upper limit on the production cross 
section of doubly charged higgs boson predicted by Left-Right symmetric
models.
\label{fig:charged}}
\end{figure}

The doubly charged higgs $H^{++}$ appears in the Left-Right symmetric 
models, SUSY, and Little Higgs models. At the Tevatron, the main production mechanism for 
doubly charged higgs is pair production. Decay modes to leptons are
largely unrestricted including possible lepton flavor violation (LFV), while decays
to $W$'s are suppressed by the $\rho$ parameter. CDF has published a paper \cite{cdf-h++}
on a search for pair produced doubly charged higgs bosons decaying to $ee$, 
$e\mu$ and $\mu \mu$ in the context of a LR-symmetric model\cite{lr-models} 
strengthening previous limits from LEP. 
The current analysis searches for $H^{++} \to e \tau$ and $\mu \tau$, two out of the 
three remaining modes (the other one is $\tau \tau$). LEP data excludes $H^{++}$
of these kinds for $m_H>97-99$ $\gevcc$. 

Analysis starts by requiring an electron or muon candidate with $p_T>20$ $\gevc$
and a hadronic tau with $p_T>15$ $\gevc$. At least one additional  
``isolated track system'' (ITS) is required (in $e+\tau$ analysis this system has
to additionally match a calorimeter cluster). All passing events are 
divided into 3- and 4-particle categories: $l+\tau+ITS$ and $l+\tau+ITS+ITS$. 
A set of event topology cuts are applied to suppress 
remaining backgrounds in events of the 3-particle topology (4-particle
events are already very clean). The choice of the event topology cuts 
is optimized for each of the two categories individually to maximize
significance. After all cuts, the expected number of events from SM backgrounds
is a fraction of an event in each category and no events in data are found. With
no excess, the analysis sets a limit on LFV $H^{++}$ species in these channels,
as shown in Fig. \ref{fig:charged}.


\section*{Acknowledgments}
The author would like to thank the US National Science Foundation for providing partial funding 
support (Award No.: PHY-0611671).

\end{document}